\documentclass[twocolumn,prl,superscriptaddress,preprintnumbers,amsmath,amssymb]{revtex4}
\usepackage[dvips]{graphicx}
\usepackage{dcolumn}
\usepackage{bm}
\begin{document}
\title{Controlling orbital moment and spin orientation in CoO layers by strain}

\author{S. I. Csiszar}
  \affiliation{ Materials Science Center, University of Groningen,
   Nijenborgh 4, 9747 AG Groningen, The Netherlands}
\author{M. W. Haverkort}
  \affiliation{ II. Physikalisches Institut, Universit\"{a}t zu K\"{o}ln,
   Z\"{u}lpicher Str. 77, 50937 K\"{o}ln, Germany}
\author{Z. Hu}
\affiliation{ II. Physikalisches Institut, Universit\"{a}t zu K\"{o}ln,
   Z\"{u}lpicher Str. 77, 50937 K\"{o}ln, Germany}
\author{A. Tanaka}
  \affiliation{ Department of Quantum Matter, ADSM, Hiroshima University,
   Higashi-Hiroshima 739-8530, Japan}
\author{H. H. Hsieh}
  \affiliation{Chung Cheng Institute of Technology,
   National Defense University, Taoyuan 335, Taiwan}
\author{H.-J. Lin}
  \affiliation{National Synchrotron Radiation Research Center,
   101 Hsin-Ann Road, Hsinchu 30077, Taiwan}
\author{C. T. Chen}
  \affiliation{National Synchrotron Radiation Research Center,
   101 Hsin-Ann Road, Hsinchu 30077, Taiwan}
\author{T. Hibma}
  \affiliation{ Materials Science Center, University of Groningen,
   Nijenborgh 4, 9747 AG Groningen, The Netherlands}
\author{L. H. Tjeng}
\affiliation{ II. Physikalisches Institut, Universit\"{a}t zu K\"{o}ln,
  Z\"{u}lpicher Str. 77, 50937 K\"{o}ln, Germany}

\date{\today}

\begin{abstract}
We have observed that CoO films grown on different substrates show dramatic
differences in their magnetic properties. Using polarization dependent x-ray
absorption spectroscopy at the Co $L_{2,3}$ edges, we revealed that the
magnitude and orientation of the magnetic moments strongly depend on the strain
in the films induced by the substrate. We presented a quantitative model to
explain how strain together with the spin-orbit interaction determine the $3d$
orbital occupation, the magnetic anisotropy, as well as the spin and orbital
contributions to the magnetic moments. Control over the sign and direction of
the strain may therefore open new opportunities for applications in the field
of exchange bias in multilayered magnetic films.

\end{abstract}

\pacs{75.25.+z, 75.70.-i, 71.70.-d, 78.70.Dm}

\maketitle The discovery of the exchange bias phenomenon in surface-oxidized
cobalt particles about 50 years ago \cite{Meiklejohn56} marks the beginning of
a new research field in magnetism. Since then several combinations of
antiferromagnetic (AFM) and ferromagnetic (FM) thin film materials have been
fabricated and investigated \cite{Nogues99,Berkowitz99}, motivated by the
potential for applications in information technology. Numerous theoretical
\cite{Malozemoff87,Mauri87,Koon97,Schulthess98,Stiles99} and experimental
\cite{Takano97,Zaag00,Ohldag01a,Ohldag03,Camarero03,Scholl04} studies have been
devoted to unravel the mechanism(s) responsible for exchange biasing. However,
no conclusive picture has emerged yet. A major part of the problem lies in the
fact that there is insufficient information available concerning the atomic and
magnetic structure of the crucial interface between the AFM and FM material.
The important issue of, for instance, spin reorientations in the AFM films
close to the interface is hardly considered
\cite{Carey93,Ijiri98,Borchers98,Ohlsdag01b,Radu03}, and the role of epitaxial
strain herein has not been discussed at all.

In this paper we study the magnetic properties of CoO thin films epitaxially
grown on MnO(100) and on Ag(100), as model systems for an AFM material under
either tensile or compressive in-plane stress. Our objective is to establish
how the magnetic anisotropy as well as the spin and orbital contributions to
the magnetic moments depend on the lowering of the local crystal field symmetry
by epitaxial strain. Using polarization dependent x-ray absorption spectroscopy
(XAS) at the Co $L_{2,3}$ ($2p \rightarrow 3d$) edges, we observe that the
magnitude and orientation of the magnetic moments in the CoO/MnO(100) system
are very different from those in the CoO/Ag(100). We present a quantitative
model to calculate how local crystal fields together with the spin-orbit
interaction determine the magnetic properties.

The actual compositions of the CoO/MnO(100) and CoO/Ag(100) systems are
(14\AA)MnO/(10\AA)CoO/ (100\AA)MnO/Ag(001) and (90\AA)CoO/Ag(001),
respectively. The two samples were grown by molecular beam epitaxy (MBE),
evaporating elemental Mn and Co from alumina crucibles in a pure oxygen
atmosphere of $10^{-7}$ to $10^{-6}$ mbar. The base pressure of the MBE system
is in the low $10^{-10}$ mbar range. The thickness and epitaxial quality of the
films are monitored by reflection high energy electron diffraction
measurements. With the lattice constant of bulk Ag (4.09\AA) being smaller than
that of bulk CoO (4.26\AA) and MnO (4.444\AA), we find from x-ray diffraction
that CoO on Ag is slightly compressed in-plane ($a_{\parallel}\approx4.235$\AA,
$a_{\perp}\approx4.285$\AA), and from reflection high energy electron
diffraction (RHEED) that CoO sandwiched by MnO is about 4\% expanded in-plane
($a_{\parallel}\approx 4.424$\AA). The sandwich structure was used to maximize
the CoO thickness with full in-plane strain. Details about the growth will be
published elsewhere \cite{Csiszar04}.

\begin{figure}[h]
     \includegraphics[width=0.40\textwidth]{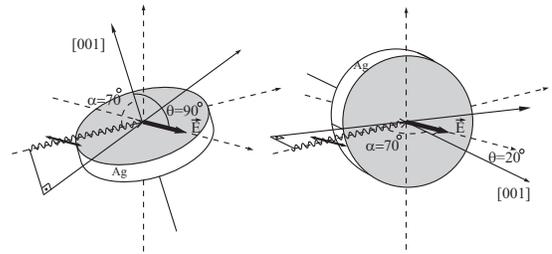}
     \caption{Experimental XAS geometry, with polarization
     of the light in the horizontal plane. $\theta$ is the angle
     between the electric field vector $\vec{E}$ and the [001] surface
     normal, and $\alpha$ the tilt between the Poynting vector and the
     surface normal.}
     \label{incidence}
\end{figure}

The XAS measurements were performed at the Dragon beamline of the NSRRC in
Taiwan using \textit{in-situ} MBE grown samples. The spectra were recorded
using the total electron yield method in a chamber with a base pressure of
3x10$^{-10}$ mbar. The photon energy resolution at the Co $L_{2,3}$ edges
($h\nu \approx 770-800$ eV) was set at 0.3 eV, and the degree of linear
polarization was $\approx 98 \%$. The sample was tilted with respect to the
incoming beam, so that the Poynting vector of the light makes an angle of
$\alpha=70^{\circ}$ with respect to the [001] surface normal. To change the
polarization, the sample was rotated around the Poynting vector axis as
depicted in Fig. 1, and $\theta$, the angle between the electric field vector
$\vec{E}$ and the [001] surface normal, can be varied between $20^{\circ}$ and
$90^{\circ}$. This measurement geometry allows for an optical path of the
incoming beam which is independent of $\theta$, guaranteeing a reliable
comparison of the spectral line shapes as a function of $\theta$. A CoO single
crystal is measured \textit{simultaneously} in a separate chamber to obtain a
relative energy reference with an accuracy of better than 0.02 eV.

\begin{figure*}
     \includegraphics[width=0.80\textwidth]{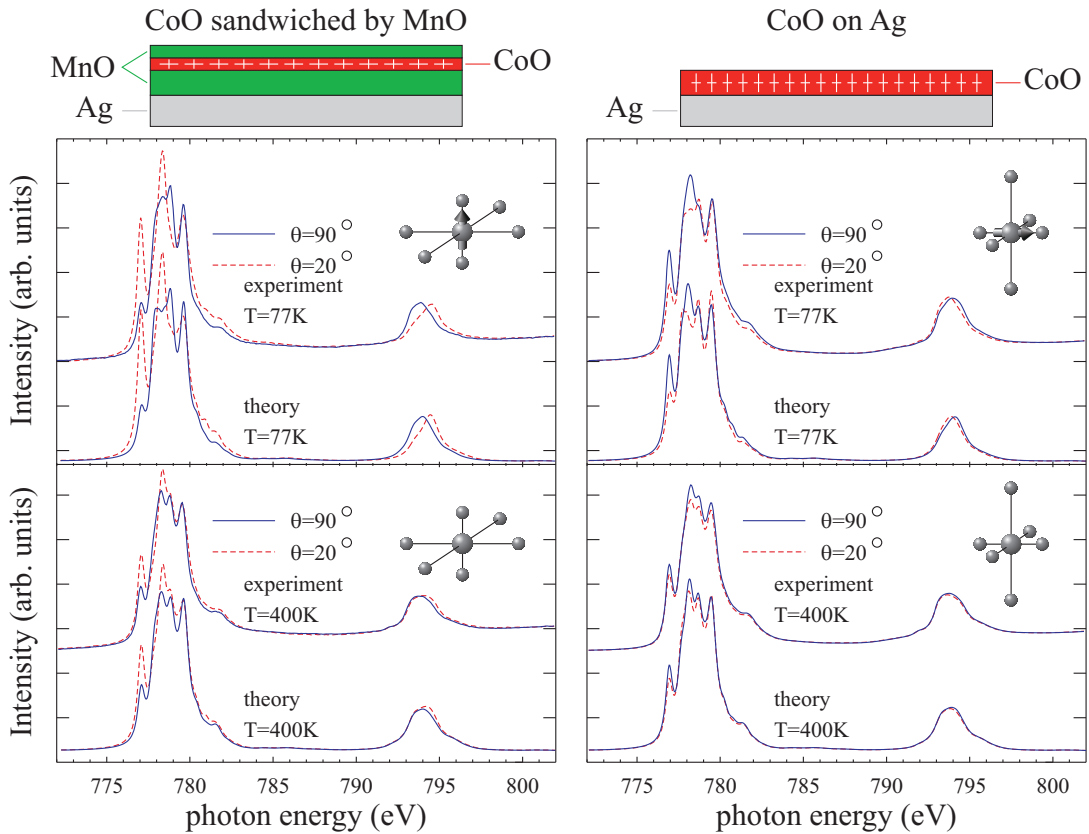}
     \caption{(color online) Experimental and calculated Co $L_{2,3}$ XAS spectra
     of: left panel) CoO in (14\AA)MnO/(10\AA)CoO/(100\AA)MnO/Ag(001) at
     $\theta=20^{\circ}$ and $\theta=90^{\circ}$, far below (top panel, T=77K)
     and far above (bottom panel, T=400K) the N\'{e}el temperature of the
     CoO thin film;
     right panel) the same for CoO in (90\AA)CoO/Ag(001)}
    \label{spectra}
\end{figure*}

Fig. 2 shows the polarization dependent Co $L_{2,3}$ XAS spectra of the
CoO/MnO(100) (left panels) and CoO/Ag(100) (right panels) systems, taken at
temperatures both far below (top panels) and far above (bottom panels) the
N\'{e}el temperature of the CoO thin film, which is about 290 K for
CoO/MnO(100) and 310 K for CoO/Ag(100) as we will show below. The angle
$\theta$ between the electric field vector $\vec{E}$ and the [001] surface
normal, is varied between $20^{\circ}$ and $90^{\circ}$. The spectra have been
corrected for electron yield saturation effects \cite{Nakajima99}. The general
line shape of the spectra shows the characteristic features similar to that of
bulk CoO \cite{deGroot94}, ensuring the good quality of our CoO films. Very
striking in the spectra is the clear polarization dependence, which is stronger
at 77K than at 400 K. Important is also that the polarization dependence of the
CoO/MnO(100) system is opposite to that of the CoO/Ag(100): for instance, the
intensity of the first peak at $h\nu$ = 777 eV is always higher for
$\theta=20^{\circ}$ than for $\theta=90^{\circ}$ in CoO/MnO(100), while it is
always smaller in CoO/Ag(100).

\begin{figure}
\includegraphics[width=0.40\textwidth]{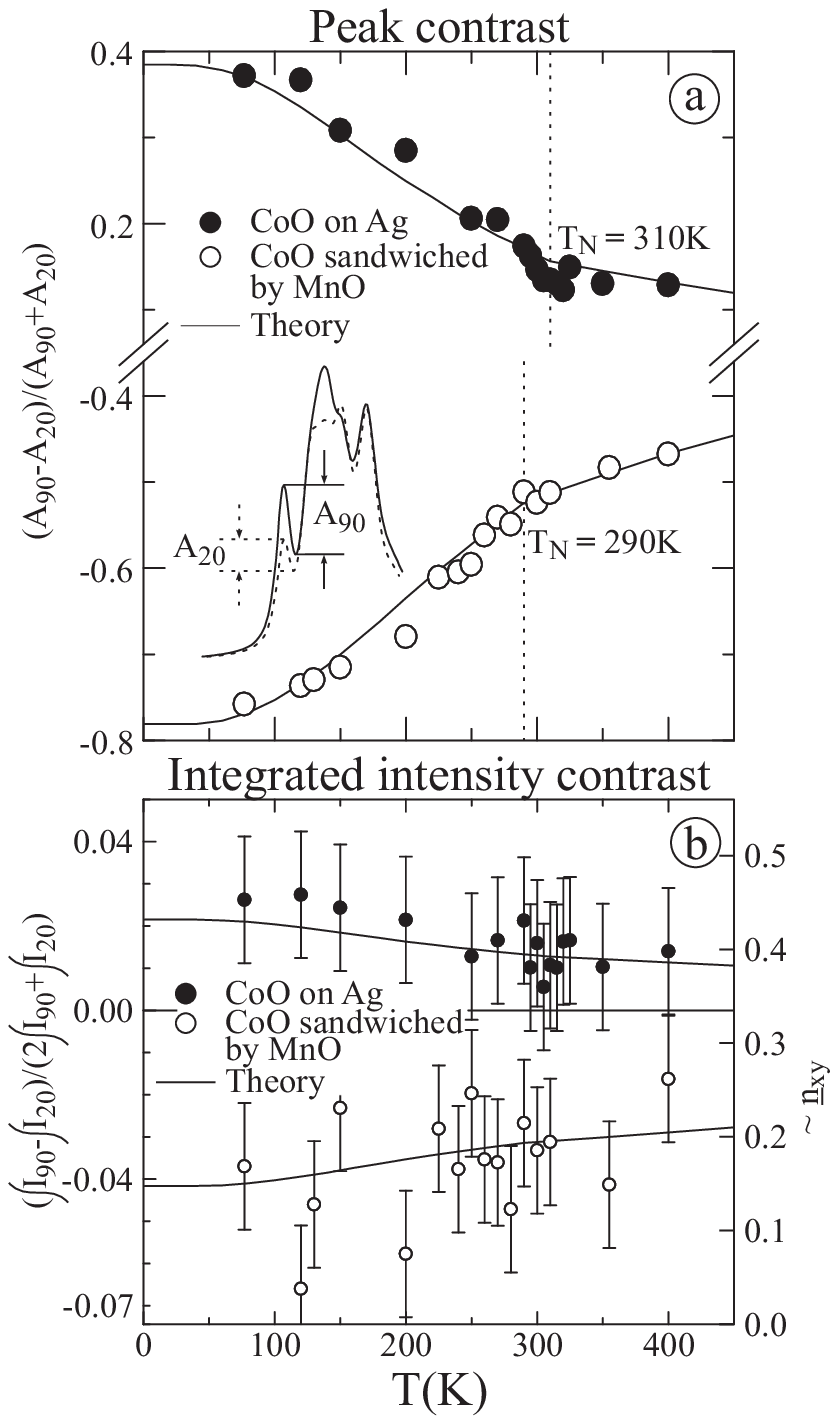}
     \caption{Temperature dependence of the polarization contrast
     in the Co $L_{2,3}$ spectra:
     a) peak contrast, defined as the difference divided by the sum
     of the height of the first peak at $h\nu$ = 777 eV, taken with
     $\theta=20^{\circ}$ and $\theta=90^{\circ}$ polarizations;
     b) integrated intensity contrast, defined as the difference divided
     by the sum of the intensity, integrated over the entire
     $L_{2,3}$ range. Filled and empty
     circles are the experimental data. The solid lines are the
     theoretical simulations.}
     \label{branching}
   \end{figure}

In order to resolve the origin of the polarization dependence in the CoO
spectra, we have investigated the temperature dependence in more detail. Fig.
3a depicts the polarization contrast of the peak at $h\nu$ = 777 eV, defined as
the difference divided by the sum of the peak height in the spectra taken with
the $\theta=20^{\circ}$ and $\theta=90^{\circ}$ polarizations. One can clearly
see a significant temperature dependence for both systems, and a closer look
also reveals the presence of a kink at about 290 K for CoO/MnO(100) and 310 K
for CoO/Ag(100), which can be associated with the N\'{e}el temperatures of the
CoO thin films as we will show below. We therefore infer that at high
temperatures the polarization contrast are caused solely by crystal field
effects, and that at low temperatures the magnetism must also contribute to the
contrast. In other words, we are observing crystal field induced linear
dichroism at high temperatures and a combination of crystal field induced and
magnetic linear dichroism at low temperatures. Important is to note that the
opposite sign in the dichroism for the CoO/MnO(100) and CoO/Ag(100) systems
implies that the orientation of the magnetic moments is perpendicular and the
sign of the crystal field splittings is opposite in the two systems.

To analyze the Co $L_{2,3}$ spectra quantitatively, we perform calculations for
the atomic-like $2p^{6}3d^{7} \rightarrow 2p^{5}3d^{8}$ transitions using a
similar method as described by Alders \textit{et al.} \cite{Alders98}, but now
in a $D_{4h}$ point group symmetry and including covalency. The method uses a
CoO$_6$ cluster which includes the full atomic multiplet theory and the local
effects of the solid \cite{deGroot94,Tanaka94}. It accounts for the
intra-atomic $3d$-$3d$ and $2p$-$3d$ Coulomb and exchange interactions, the
atomic $2p$ and $3d$ spin-orbit couplings, the O $2p$ - Co $3d$ hybridization,
local crystal field parameters $10Dq$, $Ds$ and $Dt$, and a Brillouin type
temperature dependent exchange field which acts on spins only and which
vanishes at T$_N$. The calculations have been carried out using the XTLS 8.0
programm\cite{Tanaka94}.

The results of the calculations are shown in Fig. 2. We have used the
parameters already known for bulk CoO \cite{Tanaka94,parameters}, and have to
tune only the parameters for $Ds$, $Dt$ and the direction of the exchange
field. For the CoO/MnO(100) system we find an excellent simulation of the
experimental spectra for $Ds$ = -40 meV, $Dt$ = -13 meV and an exchange field
parallel to the [001] surface normal. For the CoO/Ag(100) system we obtained
the best fit for $Ds$ = 13 meV, $Dt$ = 4 meV and an exchange field
perpendicular to the [001] surface normal. These two sets of parameters
reproduce extremely well the spectra at all temperatures, as is also
demonstrated in Fig. 3a, showing the excellent agreement between the calculated
and measured temperature dependent polarization contrast of the first peak.
Most important is obviously the information that can be extracted from these
simulations. We find that the magnetic moments in CoO/MnO(100) are oriented
out-of-plane, and, in strong contrast, those in CoO/Ag(100) to be in-plane. We
also find for CoO/MnO(100) a 2.46 $\mu_B$ spin and 1.36 $\mu_B$ orbital
contribution to the 3.82 $\mu_B$ total magnetic moment. By comparison, the
numbers for CoO/Ag(100) are smaller: 2.14, 1.00, and 3.14 $\mu_B$,
respectively. The crystal field parameters give also a very different splitting
in the $t_{2g}$ levels: about -56 meV for CoO/MnO(100) and +18 meV for
CoO/Ag(100), which is fully consistent with our structural data in that the CoO
in CoO/MnO(100) experiences a \textit{large} in-plane \textit{expansion}
(tensile strain) while the CoO in CoO/Ag(100) is only \textit{slightly}
\textit{contracted} in-plane (compressive strain).

Shape anisotropy cannot explain why the spin of the thinner CoO film, i.e.
CoO/MnO(100) is oriented out-of-plane while that of the thicker film, i.e.
CoO/Ag(100), is in-plane. In order to understand intuitively the important
interplay between strain and spin-orbit interaction for the magnetic properties
of materials with a partially filled $3d$ $t_{2g}$ shell, we will start with
describing the energetics of the high spin Co$^{2+}$ ($3d^{7}$) ion in a
one-electron like picture. In $O_{h}$ symmetry the atomic $3d$ levels are split
into 3 $t_{2g}$ and 2 $e_{g}$ orbitals, so that two holes reside in the
spin-down $e_g$ orbitals and one hole in one of the spin-down $t_{2g}$.
Introducing a tetragonal distortion, in which the c-axis (out-of-plane) is made
different from the a-axis (in-plane), the $t_{2g}$ levels also become split. In
the limit that this splitting is much larger than the spin-orbit interaction,
we will find for CoO with c/a$\ll$1, that the $t_{2g}$ hole will occupy a
linear combination of ${d}_{xz}$ and ${d}_{yz}$ orbitals. The spin-orbit
interaction will then produce a $m_l = -1$ state, i.e. a state with an orbital
moment of 1 $\mu_{B}$ directed perpendicular to the plane of the film. The spin
moment will be also out-of-plane, since it is coupled via the spin-orbit
interaction to the orbital moment. For CoO with c/a$\gg$1, we will get a
$t_{2g}$ hole in the $d_{xy}$ orbital, i.e. a state with a quenched orbital
momentum \cite{Jo98}. For the actual CoO/Ag(100) system, we find that c/a is
indeed larger than 1, but only slightly and with a $t_{2g}$ splitting which is
smaller than the spin-orbit interaction. As a result, the orbital moment is not
quenched \cite{Haverkort04}. In fact, it is directed in-plane, and thus also
the spin moment. For the CoO/MnO(100) system, the c/a is smaller than 1, and
the orbital and spin moment are indeed directed out-of-plane. The size of the
orbital moment as calculated in the cluster model is somewhat larger than 1
$\mu_{B}$, but this is merely due to the presence of Coulomb and exchange
interactions in the multiplet structure \cite{Haverkort04,Sugano70} not
considered in the simple one-electron picture.

The fact that strain has a direct influence on the orbital occupation as
revealed by our theoretical analysis, can also be verified experimentally. In
Fig. 3b we plot the difference divided by the sum of the integrated Co
$L_{2,3}$ intensities for the two polarizations for a wide range of
temperatures. The integration is over the entire Co $L_{2,3}$ spectral region
and a background has been subtracted following the prescription often used in
evaluating orbital moment sum rules in soft-x-ray magnetic circular dichroism
spectroscopy \cite{Chen95}. One can clearly observe that there is a significant
non-zero polarization contrast in the integrated intensities, and that this
contrast has an opposite sign for the two systems studied here. It is important
to realize that the integrated intensity for a particular polarization of the
light merely depends on the symmetry of the $3d$ holes in the initial state. To
illustrate this, let us make the simplification that 2 holes reside in the
spin-down $e_g$ orbitals and 1 hole in one of the spin-down $t_{2g}$. We then
have \cite{Haverkort04}:
\begin{equation}
\frac{\int I_{90}-\int I_{0}}{2\int I_{90}+\int I_{0}} =
\frac{\underline{n}_{xy}-\frac{1}{2}(\underline{n}_{xz}+\underline{n}_{yz})}{2\underline{n}_{total}}
\end{equation}
where $\underline{n}$ denotes the hole number and the subscript the orbital
type. For an isotropic orbital occupation, the integrated contrast would be 0
and $\underline{n}_{xy}$ = 1/3. One can now directly deduce from the data that
the CoO/Ag(100) system has relatively more $d_{xy}$ holes than the
CoO/MnO(100), fully consistent with $c/a>1$ in CoO/Ag(100), and $c/a<1$ in
CoO/MnO(100). The cluster calculations reproduce the measured integrated
polarization contrast very well, and reveal that the actual
$\underline{n}_{xy}$ \cite{Haverkort04} is about 0.4 and 0.2 for the two
systems.

Based on the parameters extracted from the excellent simulations of the
spectra, we have estimated the magnetocrystalline anisotropy within the CoO
films. We calculated the single ion anisotropy by comparing the total energy of
the CoO$_6$ cluster for different exchange field directions. This energy is
expressed as $E = K_{0} + K_{1}sin^{2}(\eta) + K_{2}sin^{4}(\eta) +
K_{3}sin^{4}(\eta)sin^{2}(\varphi)cos^{2}(\varphi)$, where $\eta$ is the angle
between the exchange field and the c-axis, and $\varphi$ the azimuthal angle.
We find for CoO/MnO(100) $K_{1}$ = 3.4 meV, $K_{2}$ = 1.4 meV and $K_{3}$ = 0.1
meV, while for CoO/Ag(100) we obtain $K_{1}$ = -1.7 meV, $K_{2}$ = 0.1 meV, and
$K_{3}$ = 0.1 meV. In other words, for CoO/MnO(100) the energy difference
between the spin directed parallel with the easy axis ($\eta = 0^{\circ}$,
perpendicular to the film) and parallel with the hard axis ($\eta =
90^{\circ}$, lying in the film plane) to be about 4.8 meV. The same single ion
anisotropy energy calculated for the CoO/Ag(100) is about -1.6 meV, i.e. the
hard axis is perpendicular and the easy axis parallel to the film plane. These
energies are more than 2 orders of magnitude larger than the dipolar anisotropy
within the film \cite{Finazzi03}.

To conclude, CoO films grown on different substrates show dramatic differences
in their magnetic properties. Strain induced local crystal fields together with
the spin-orbit interaction determine the $3d$ orbital occupation, the magnetic
anisotropy, as well as the spin and orbital contributions to the magnetic
moments. Control over the sign and direction of the strain may therefore open
new opportunities for applications in the field of exchange bias in
multilayered magnetic films, especially when using magnetic ions with a
partially field $t_{2g}$ shell.

We acknowledge the NSRRC staff for providing us with an extremely stable beam.
We would like to thank Lucie Hamdan and Henk Bruinenberg for their skillful
technical and organizational assistance in preparing the experiment. The
research in Cologne is supported by the Deutsche Forschungsgemeinschaft through
SFB 608.

\end{document}